\documentclass[%
 aip,
 jmp,%
 amsmath,amssymb,
reprint,%
]{revtex4-1}

\usepackage{graphicx}
\usepackage{dcolumn}
\usepackage{bm}
\usepackage{subfigure} 

\begin{document}


\title[Nucleation rate and flux of composite nucleus]{Steady-state nucleation rate and flux of composite nucleus at saddle point 
}

\author{Masao Iwamatsu}
\email{iwamatsu@ph.ns.tcu.ac.jp}
\affiliation{ 
Department of Physics, Faculty of Liberal Arts and Sciences, Tokyo City University, Setagaya-ku, Tokyo 158-8557, JAPAN
}%

\date{\today}

\begin{abstract}
The steady-state nucleation rate and flux of composite nucleus at the saddle point is studied by extending the theory of binary nucleation.  The Fokker-Planck equation that describes the nucleation flux is derived using the Master equation for the growth of the composite nucleus, which consists of the core of the final stable phase surrounded by a wetting layer of the intermediate metastable phase nucleated from a metastable parent phase recently evaluated by the author [J. Chem. Phys. {\bf 134}, 164508 (2011)].  The Fokker-Planck equation is similar to that used in the theory of binary nucleation, but the non-diagonal elements exist in the reaction rate matrix.  First, the general solution for the steady-state nucleation rate and the direction of nucleation flux is derived. Next, this information is then used to study the nucleation of composite nucleus at the saddle point. The dependence of steady-state nucleation rate as well as  the direction of nucleation flux on the reaction rate in addition to the free-energy surface is studied using a model free-energy surface.  The direction of nucleation current deviates from the steepest-descent direction of the free-energy surface.  The results show the importance of two reaction rate constants: one from the metastable environment to the intermediate metastable phase and the other from the metastable intermediate phase to the stable new phase.  On the other hand, the gradient of the potential $\Phi$ or the Kramers crossover function (the commitment or splitting probability) is relatively insensitive to reaction rates or free-energy surface.
\end{abstract}

\maketitle

\section{\label{sec:sce1}Introduction}
Nucleation is a very basic phenomena which plays a vital role in various material processing in industry ranging from steel production to food and beverage industries~\cite{Kelton2010}.  Recently, researchers have focused on the nucleation of complex materials which are relevant to our daily life~\cite{Kelton2010}.  The nucleation of such complex materials can also be complex and may involve intermediate metastable phases~\cite{Ostwald1897,Vekilov2010, Gebauer2008,Chung2009}.  These metastable intermediate phases indicate a composite nucleus that consists of a core of stable new phase surrounded by a wetting layer~\cite{Vekilov2010} of intermediate phase, which is nucleated from the metastable parent phase.

The author~\cite{Iwamatsu2011} recently examined the thermodynamics and free-energy surface of a critical nucleus using capillarity approximation when an intermediate metastable phase exists  The author discovered the possibility of such a composite critical nucleus at the saddle point of the free-energy surface~\cite{Iwamatsu2011}. In order to study the free-energy surface of nucleation, we utilized a three-phase system and assumed the existence of an intermediate metastable phase in addition to an unstable parent phase and a stable new phase.  We chose a composite binary nucleus that consists of the stable new phase and the metastable intermediate phase.  This binary nucleus is completely phase separated, and the metastable phase is segregated on to the surface.  By studying the free-energy surface of the two-component system in the two-dimensional space~\cite{Iwamatsu2011}, we can easily visualize the free-energy surface and are able to locate the saddle point that corresponds to the critical nucleus.  Even though such a composite nucleus has long been predicted theoretically~\cite{Kashchiev1998,Kashchiev2005,Meel2008}, it has been proved explicitly that the composite nucleus corresponds to the saddle point only through the use of a computer simulation~\cite{tenWolde1997}.   Our calculations~\cite{Iwamatsu2011} clearly demonstrate that by using an analytical model the composite nucleus really corresponds to the saddle point of the free-energy surface.  A similar composite nucleus at the saddle point of the free-energy surface is considered in the problem of deliquescence~\cite{Djikaev2002,Shchekin2008,McGraw2009}.

From the free-energy surface, the author could easily calculate the so-called minimum-free-energy path (MFEP)~\cite{Iwamatsu2009}.  However, it has been recognized for more than thirty years that the steepest-descent direction on the free-energy surface~\cite{Reiss1950} that corresponds to the MFEP does not necessarily indicate the real nucleation pathway in binary nucleation~\cite{Stauffer1976,Temkin1984,Greer1990}.  This is because the asymmetry of the reaction rate will deflect the direction of nucleation flux from the steepest-descent direction. Likewise, the free-energy surface alone may not determine the nucleation pathway for the composite nucleus.  The reason for this is that the asymmetry between the reaction rate from the metastable parent phase to the intermediate metastable phase and that from the intermediate to stable phase will play some role to determine not only the nucleation rate but the direction of nucleation flux.  Therefore, the MFEP determined by the steepest-descent direction of free-energy surface~\cite{Iwamatsu2009} will not necessarily correspond to the {\it real} nucleation pathway.

This study focuses on the steady-state nucleation rate and flux at the saddle point for the composite nucleus through the theory of binary nucleation~\cite{Trinkaus1983,Wilemski1999}.  We must first determine the Fokker-Planck or the Zeldovich-Frenkel equation~\cite{Kelton2010,Wu1997,Zeldovich1943,Frenkel1955} from the Master equation for the composite nucleus (Section \ref{sec:sec2}).  The  theory of nucleation rate and flux for the binary nucleation when the reaction rate matrix is non-diagonal is re-constructed in Section~\ref{sec:sec3} and in the Appendix A and B as the theory of Trinkaus~\cite{Trinkaus1983} is too condensed to study various aspects of nucleation of multicomponent systems.  In section \ref{sec:sec4} the theory developed in Sections \ref{sec:sec2} and \ref{sec:sec3} will be applied to the composite nucleus.  An example to illustrate the process will be provided in order to show how the reaction rate will influence the nucleation rate and the direction of nucleation flux at the saddle point for the composite nucleus.  Section \ref{sec:sec5} will contain the conclusion of the study.

\section{\label{sec:sec2}Fokker-Planck equation for the composite nucleus}

\begin{figure}[htbp]
\begin{center}
\includegraphics[width=0.65\linewidth]{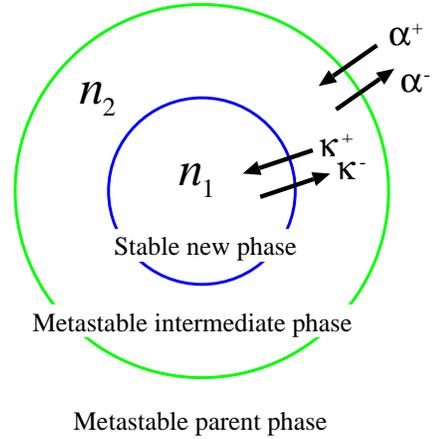}
\end{center}
\caption{
A composite critical nucleus model that consists of a stable new phase (number of molecules $n_1$) surrounded by an intermediate metastable phase (number of molecules $n_2$) nucleated in the metastable parent phase~\cite{Iwamatsu2011}.  Nucleation rates $\kappa^{+}$ and $\kappa^{-}$ are the reaction rates between the stable new phase and the metastable intermediate phase.  Nucleation rates $\alpha^{+}$ and $\alpha^{-}$ are the reaction rate or attachment rate from the metastable parent phase to the intermediate metastable phase.   } 
\label{fig:1x}
\end{figure}

In order to study the nucleation kinetics of the composite nucleus, the model shown in Fig.~\ref{fig:1x} is considered.  The model consists of a core of the stable new phase (number of molecules $n_1$) surrounded by an intermediate metastable phase (number of molecules $n_2$) nucleated in the metastable parent phase~\cite{tenWolde1997, Meel2008, Vekilov2010, Iwamatsu2011}.  Nucleation rates $\kappa^{+}$ and $\kappa^{-}$ are the reaction rates between the stable new phase and the metastable intermediate phase.  Nucleation rates $\alpha^{+}$ and $\alpha^{-}$ are the reaction rate or attachment rate from the metastable parent phase to the intermediate metastable phase.  

A similar model has been used to study the partitioning transformation~\cite{Russell1968,Kelton2000,Diao2008} where the coupling of the interfacial and long-range diffusion fluxes is considered to study the time-dependent precipitate nucleation~\cite{Kelton2003}, void formation in irradiated metals~\cite{Russell1978}, and nanocrystal formation from metallic glasses~\cite{Kelton1998}.  In these linked-flux models~\cite{Russell1968,Kelton2000,Diao2008}, in contrast to our model~\cite{Iwamatsu2011}, the metastable intermediate phase does not really represent the thermodynamics phase.  Rather, it models a transition zone due to the diffusion of monomers~\cite{Russell1968, Kelton2000}.  Therefore, the nucleation rate $\alpha^{+}$ does not represents the attachment rate of monomer but is simply a parameter which approximately represents diffusion.  In our model, however, the intermediate phase represents the metastable thermodynamics phase and $\alpha^{+}$ is the attachment rate of the monomer.  

The Master equation for the time-dependence of the number of clusters $f\left(n_1,n_2,t\right)$ that consists of $n_1$ molecules of the stable phase in the core and $n_2$ molecules of the intermediate metastable phase  in the surrounding wetting layer is written generally~\cite{Kelton2010,Wu1997, Reiss1950} in the form
\begin{equation}
\frac{\partial f\left(n_1,n_2,t\right)}{\partial t}=\left(-J_{n_1}+J_{n_1-1}\right)
+\left(-J_{n_2}+J_{n_2-1}\right),
\label{eq:2z}
\end{equation}
where
\begin{eqnarray}
J_{n_1}&=&\kappa^{+}\left(n_1,n_2\right)f\left(n_1,n_2,t\right) \nonumber \\
&-&\kappa^{-}\left(n_1+1,n_2-1\right)f\left(n_1+1,n_2-1,t\right), \nonumber \\
J_{n_1-1}&=&\kappa^{+}\left(n_1-1,n_2+1\right)f\left(n_1-1,n_2+1,t\right) \nonumber \\
&-&\kappa^{-}\left(n_1,n_2\right)f\left(n_1,n_2,t\right), 
\label{eq:3z} \\
J_{n_2}&=&\alpha^{+}\left(n_1,n_2\right)f\left(n_1,n_2,t\right) \nonumber \\
&-&\alpha^{-}\left(n_1,n_2+1\right)f\left(n_1,n_2+1,t\right), \nonumber \\
J_{n_2-1}&=&\alpha^{+}\left(n_1,n_2-1\right)f\left(n_1,n_2-1,t\right) \nonumber \\
&-&\alpha^{-}\left(n_1,n_2\right)f\left(n_1,n_2,t\right). \nonumber
\end{eqnarray}
Using the detailed balance condition, 
\begin{eqnarray}
\kappa^{-}\left(n_1+1,n_2-1\right)&=&\kappa^{+}\left(n_1,n_2\right)\frac{f_{\rm eq}\left(n_1,n_2\right)}{f_{\rm eq}\left(n_1+1,n_2-1\right)}, \nonumber \\
\alpha^{-}\left(n_1,n_2+1\right)&=&\alpha^{+}\left(n_1,n_2\right)\frac{f_{\rm eq}\left(n_1,n_2\right)}{f_{\rm eq}\left(n_1,n_2+1\right)},
\label{eq:4z}
\end{eqnarray}
where the equilibrium cluster distribution $f_{\rm eq}\left({\bm n}\right)$ is given by the usual Boltzmann distribution
\begin{equation}
f_{\rm eq}\left({\bm n}\right)=f_{0}\exp\left(-\beta G\left({\bm n}\right)\right),
\label{eq:5z}
\end{equation}
and $G\left({\bm n}\right)$ is the work of cluster formation for a cluster with composition ${\bm n}=\left(n_1,n_2\right)$ and $\beta$ is the inverse temperature.   Then, Eq.~(\ref{eq:2z}) can be written as
\begin{eqnarray}
&&\frac{\partial f\left(n_1,n_2,t\right)}{\partial t}= \nonumber \\
&-&\kappa^{+}\left(n_1,n_2\right)f_{\rm eq}\left(n_1,n_2\right) \nonumber \\ 
&&\times\left[\frac{f\left(n_1,n_2,t\right)}{f_{\rm eq}\left(n_1,n_2\right)}-\frac{f\left(n_1+1,n_2-1,t\right)}{f_{\rm eq}\left(n_1+1,n_2-1\right)}\right]
\nonumber \\
&+&\kappa^{+}\left(n_1-1,n_2+1\right)f_{\rm eq}\left(n_1-1,n_2+1\right) \nonumber \\
&&\times\left[\frac{f\left(n_1-1,n_2+1,t\right)}{f_{\rm eq}\left(n_1-1,n_2+1\right)} 
-\frac{f\left(n_1,n_2,t\right)}{f_{\rm eq}\left(n_1,n_2\right)}\right] \nonumber \\
&-&\alpha^{+}\left(n_1,n_2\right)f_{\rm eq}\left(n_1,n_2\right) \nonumber \\
&&\times\left[\frac{f\left(n_1,n_2,t\right)}{f_{\rm eq}\left(n_1,n_2\right)} 
-\frac{f\left(n_1,n_2+1,t\right)}{f_{\rm eq}\left(n_1,n_2+1\right)}\right] \nonumber \\
&+&\alpha^{+}\left(n_1,n_2-1\right)f_{\rm eq}\left(n_1,n_2-1\right) \nonumber \\
&&\times\left[\frac{f\left(n_1,n_2-1,t\right)}{f_{\rm eq}\left(n_1,n_2-1\right)} 
-\frac{f\left(n_1,n_2,t\right)}{f_{\rm eq}\left(n_1,n_2\right)}\right].
\label{eq:6z}
\end{eqnarray}
In the continuum limit, we have
\begin{equation}
\frac{\partial f\left({\bm n}\right)}{\partial t}=-\left(\frac{\partial J_{n_1}}{\partial n_1}+\frac{\partial J_{n_2}}{\partial n_2}\right)=-{\rm div}{\bm J},
\label{eq:10z}
\end{equation}
where the components of the nucleation flux $\bm J$ are given by
\begin{eqnarray}
J_{n_1}&=-&\kappa^{+}f_{\rm eq}\left\{\frac{\partial \Phi}{\partial n_1}
-\frac{\partial \Phi}{\partial n_2}\right\},  \label{eq:11z}
\\
J_{n_2}&=&-\alpha^{+} f_{\rm eq}\left\{\frac{\partial \Phi}{\partial n_2}\right\}  \nonumber \\
&&-\kappa^{+}f_{\rm eq}\left\{-\frac{\partial \Phi}{\partial n_1}
+\frac{\partial \Phi}{\partial n_2}\right\},
\label{eq:12z}
\end{eqnarray}
and
\begin{equation}
\Phi\left(n_1,n_2,t\right)=\frac{f\left(n_1,n_2,t\right)}{f_{\rm eq}\left(n_1,n_2\right)},
\label{eq:9z}
\end{equation}
and simplified notations $\kappa^{+}=\kappa^{+}\left(n_1,n_2\right)$, $\alpha^{+}=\alpha^{+}\left(n_1,n_2\right)$ and $f_{\rm eq}=f_{\rm eq}\left(n_1,n_2\right)$ and $\Phi=\Phi\left(n_1,n_2,t\right)$ are used.  Equation (\ref{eq:10z}) is called the Fokker-Planck~\cite{Risken1989} or the Zeldovich-Frenkel equation~\cite{Zeldovich1943,Frenkel1955}.

Equations (\ref{eq:11z}) and (\ref{eq:12z}) can be put in the form of a matrix equation 
\begin{equation}
{\bm J}_{\bm n}=-f_{\rm eq}\left({\bm n}\right){\bm R\left({\bm n}\right)}{\bm \nabla}_{\bm n}\Phi\left({\bm n}\right),
\label{eq:15z}
\end{equation}
where $\bm J$ and ${\bm \nabla}_{\bm n}$ are row vectors, and $\bm R$ is a symmetric square matrix defined through Eqs.~(\ref{eq:11z}) and (\ref{eq:12z}).  The two fluxes $J_{n_1}$ and $J_{n_2}$ are linked by the non-diagonal rate matrix $R$. Eq.~(\ref{eq:15z}) is originally derived by Russell~\cite{Russell1968} though the analysis of the nucleation flux at the saddle point was mostly qualitative and numerical and did not present various analytical formulas which will be derived in this paper.

A general theory of barrier crossing by the decay of metastable state has been formulated by Langer~\cite{Langer1969} using the Fokker-Planck equation more than forty years ago by generalizing the Kramers' theory~\cite{Kramers1940} of barrier crossing, and has been extended by several workers~\cite{Berezhkovskii2005, Peters2009} to study the reaction and nucleation pathways in various problems not limited to binary nucleation.  This Kramers-Langer-Berezhkovskii-Szabo (KLBS) theory~\cite{Kramers1940,Langer1969,Berezhkovskii2005} considers a general problem of escape probability from a potential well of the metastable state.  In this KLBS theory, the nucleation flux ${\bm J}$ is the probability flux and $\Phi$ is called the Kramers crossover function~\cite{Peters2009} or the commitment or splitting probability~\cite{Rhee2005} which represents the probability to cross the barrier from the given point in the phase space.

In this paper, we adopt the theory of binary nucleation~\cite{Reiss1950,Stauffer1976,Trinkaus1983,Temkin1984,Greer1990,Wu1997,Wilemski1999} since it does not need to characterize the initial metastable state by a potential well.  Also, the connections to the original Master equation are more transparent since the elementary process of nucleation is a discrete process of attachment and detachment of monomers.  It is well recognized~\cite{Stauffer1976} that we must return to the original discrete Master equation from the Fokker-Planck equation when we study the nucleation pathway of post-critical nucleus.  In any case, the formula based on KLBS theory~\cite{Kramers1940,Langer1969,Berezhkovskii2005} reduces to that used in the theory of binary nucleation~\cite{Trinkaus1983} when the thermal distribution in the metastable potential well is incorporated~\cite{Zitserman1990}.  We will extend the theory~\cite{Reiss1950,Stauffer1976,Trinkaus1983,Temkin1984,Greer1990,Wu1997,Wilemski1999} developed for binary nucleation to study the nucleation of composite nucleus in the next section.

\section{\label{sec:sec3} Magnitude and direction of steady state nucleation flux at the saddle point}

From this point, we will concentrate on the steady state nucleation flux at the saddle point ${\bm n}^{*}$ so that the nucleation flux is a time independent constant vector (${\rm div}{\bm J}=0$) and the rate matrix ${\bm R}={\bm R}\left({\bm n}^{*}\right)$ is also a constant matrix at the saddle point.

The general theory of nucleation flux of binary nucleation will be reformulated when the rate matrix $\bm R$ is non-diagonal so that we can study the nucleation of composite nucleus.   Most of the previous authors who studied the nucleation flux of binary nucleation have studied the cases where the rate matrix is diagonal~\cite{Stauffer1976,Wilemski1999}.  To exemplify this, we introduce the rotation matrix of the coordinate given by
\begin{equation}
{\bm V}\left(\theta\right)=
\begin{pmatrix}
\cos\theta & -\sin\theta \\
\sin\theta  &  \cos\theta 
\end{pmatrix},
\label{eq:17z}
\end{equation}
which is the orthogonal transformation in vector space, and study the steady-state solution $\partial f/\partial t=0$ of the Fokker-Planck equation (\ref{eq:10z}) given by
\begin{equation}
{\bm \nabla}_{\rm n}^{\rm T}{\bm J}=\partial_{n_1}J_{n_1}+\partial_{n_2}J_{n_2}=0,
\label{eq:19z}
\end{equation}
where we have used a column vector ${\bm \nabla}_{\rm n}^{\rm T}$ for the differential operator ${\rm div}$ and the superscript T which means the transpose.

Using the rotation matrix in Eq.~(\ref{eq:17z}),  we can introduce a new coordinate ${\bm e}_{\bm \eta}$ whose origin is displaced at the saddle point ${\bm n}^{*}$ and rotate the unit vectors ${\bm e}_{n_1}$ and ${\bm e}_{n_2}$ by an angle $\theta$ through
\begin{equation}
{\bm e}_{\bm n}={\bm V}\left(\theta\right){\bm e}_{\bm \eta},\;\;\;
\mbox{or}\;\;\;\;
{\bm e}_{\bm \eta}={\bm V}^{\rm T}\left(\theta\right){\bm e}_{\bm n}.
\label{eq:21z}
\end{equation}
Then the displacement (column) vector from the saddle point ${\bm n}^{*}$
\begin{equation}
\Delta {\bm n}={\bm n}-{\bm n}^{*}=
\begin{pmatrix}
\Delta n_1 \\
\Delta n_2
\end{pmatrix},
\label{eq:22z}
\end{equation}
will be transformed into the displacement $\bm \eta$ through
\begin{equation}
\Delta{\bm n}={\bm V}\left(\theta\right){\bm \eta},\;\;\;
\mbox{or}\;\;\;\;
{\bm \eta}={\bm V}^{\rm T}\left(\theta\right)\Delta{\bm n}.
\label{eq:24z}
\end{equation}
The direction of nucleation flux given by Eq.~(\ref{eq:15z}) can be studied by inspecting the appropriate rotation of coordinate.

\subsection{\label{sec:3-1}Direction of steepest-descent of free-energy surface}

The steepest-descent direction of the free-energy surface, which is the direction of minimum free-energy path (MFEP)~\cite{Iwamatsu2009}, at the saddle point can be studied by expanding the work of cluster formation $G\left({\bm n}^{*}\right)$ around the saddle point ${\bm n}^{*}$ as~\cite{Reiss1950}
\begin{equation}
G\left({\bm n}\right)=G^{*}+\frac{1}{2}\Delta{\bm n}^{\rm T}{\bm G}\Delta{\bm n}
\label{eq:25z}
\end{equation}
using the displacement (column) vector defined in Eq.~(\ref{eq:22z}), where
\begin{equation}
{\bm G}=
\begin{pmatrix}
G_{11} & G_{12} \\
G_{21} & G_{22}
\end{pmatrix},
\label{eq:26z}
\end{equation}
with
\begin{equation}
G_{ij}=\frac{\partial^{2} G}{\partial n_i \partial n_j},
\label{eq:27z}
\end{equation}
is the Hessian matrix of the free energy at the saddle point ${\bm n}^{*}$, and $G^{*}=G\left({\bm n}^{*}\right)$.  By using the rotation matrix ${\bm V}\left(\theta\right)$ defined in Eq.~(\ref{eq:17z}), 
we can diagonalize the matrix $\bm G$ and obtain the negative eigenvalue that corresponds to the steepest-descent direction as
\begin{equation}
\lambda_1 = \left(\left(G_{11}+G_{22}\right)-\sqrt{\left(G_{11}+G_{22}\right)^2-4{\rm det}{\bm G}}\right)/2,
\label{eq:30z}
\end{equation}
where ${\rm det}{\bm G}=G_{11}G_{22}-G_{12}^{2}<0$ at the saddle point.

The rotation angle $\theta$ is chosen such that $\lambda_{1}$ becomes negative and $\lambda_2$ becomes positive.  Then the rotation angle $\theta$ is given by
\begin{equation}
\tan \theta=\frac{\left(G_{22}-G_{11}\right)}{2G_{12}}
\pm\sqrt{\left(\frac{G_{22}-G_{11}}{2G_{12}}\right)^2+1},
\label{eq:32z}
\end{equation}
where $+$ sign must be chosen when $G_{12}<0$ otherwise $-$ sign must be chosen.  The steepest-descent direction is the direction of $\eta_1$ axis defined through
from Eq.~({\ref{eq:21z}) with $\theta$ given by Eq.~(\ref{eq:32z}).  This is the direction of the eigenvector of negative eigenvalue $\lambda_1$ of the matrix $\bm G$.

\subsection{\label{sec:3-2}Direction of gradient of $\Phi$}

Next, we will consider the direction of the gradient of $\Phi$, which has been studied intensively by Wilemski~\cite{Wilemski1999} and others~\cite{Wyslouzil1999,Li2000}.  This function $\Phi$ is also known as the Kramers crossover function~\cite{Peters2009} or the commitment or splitting probability~\cite{Rhee2005} in the general theory of barrier crossing, where $\Phi$ plays the role of the probability to cross the barrier.  We will extend the work of Wilemski~\cite{Wilemski1999} since we are dealing with nucleation by an attachment and detachment process.  Now, we rotate $\left(\Delta n_1,\Delta n_2\right)$ axis by $\omega$ and introduce a new coordinate $\left(\zeta_1,\zeta_2\right)$ through
\begin{equation}
{\bm \zeta} = {\bm V}^{\rm T}\left(\omega\right)\Delta_{\bm n}.
\label{eq:35z}
\end{equation}
Then, the component $J_{\zeta_1}$ and $J_{\zeta_2}$ are related to $J_{n_1}$ and $J_{n_2}$ through
\begin{equation}
{\bm J}_{\bm \zeta}={\bm V}^{\rm T}\left(\omega\right){\bm J}_{\bm n}.
\label{eq:37z}
\end{equation}
Similarly we obtain
\begin{equation}
{\bm \nabla}_{\bm \zeta}={\bm V}^{\rm T}\left(\omega\right){\bm \nabla}_{\bm n},
\label{eq:38z}
\end{equation}
which is the simplest transformation of covariant vector~\cite{Risken1989,Graham1977,Grabert1980} when the coordinate transformation is given by a linear orthogonal transformation Eq.~(\ref{eq:17z}) as the gradient of a scalar is a covariant vector.  The covariant formulation of the Fokker-Planck equation~\cite{Risken1989,Graham1977} will be useful when we consider more general coordinate transformations~\cite{Grabert1980,Fisenko2004}. 

The nucleation flux in Eq.~(\ref{eq:15z}) is now transformed into
\begin{equation}
{\bm J}_{\bm \zeta}=-f_{\rm eq}\left({\bm n}\right){\bm V}^{\rm T}\left(\omega\right){\bm R}{\bm V}\left(\omega\right){\bm\nabla}_{\bm \zeta}\Phi
\label{eq:40z}
\end{equation}
at the saddle point.

By calculating the matrix products ${\bm V}^{\rm T}{\bm R}{\bm V}$, Eq.~(\ref{eq:40z}) is written explicitly by
\begin{eqnarray}
J_{\zeta_1} &=& -f_{\rm eq}\left({\bm n}\right)\Sigma\left(\omega\right)\frac{\partial \Phi}{\partial \zeta_1}+f_{\rm eq}\left({\bm n}\right)\Xi\left(\omega\right)\frac{\partial \Phi}{\partial \zeta_2},
\label{eq:42z} \\
J_{\zeta_2} &=& f_{\rm eq}\left({\bm n}\right)\Xi\left(\omega\right)\frac{\partial \Phi}{\partial \zeta_1}-f_{\rm eq}\left({\bm n}\right)\Upsilon\left(\omega\right)\frac{\partial \Phi}{\partial \zeta_2},
\label{eq:43z}
\end{eqnarray} 
where
\begin{eqnarray}
\Sigma\left(\omega\right) &=&
R_{11}\cos^{2}\omega+R_{22}\sin^{2}\omega+2R_{12}\cos\omega\sin\omega,
\nonumber \\
\label{eq:44z} \\
\Xi\left(\omega\right) &=&
\left(R_{11}-R_{22}\right)\cos\omega\sin\omega+R_{12}\left(\sin^{2}\omega-\cos^{2}\omega\right),
\nonumber \\
\label{eq:45z} \\
\Upsilon\left(\omega\right) &=&
R_{11}\sin^{2}\omega + R_{22}\cos^{2}\omega - 2R_{12}\cos\omega\sin\omega
\nonumber \\
\label{eq:47z}
\end{eqnarray}
are the matrix elements of ${\bm V}^{\rm T}{\bm R}{\bm V}$.

We then choose the new coordinate $\zeta$ such that $\partial_{\zeta_2}\Phi=0$.  After manipulating Eqs.~(\ref{eq:42z}) and (\ref{eq:43z}) using the same procedure originally developed by Langer~\cite{Langer1969}, which is briefly summarized in Appendix A, we find
\begin{equation}
\tan\omega = s\pm\sqrt{s^2+r},
\label{eq:59z}
\end{equation}
where
\begin{eqnarray}
s &=& \frac{R_{22}G_{22}-R_{11}G_{11}}{2\left(R_{12}G_{11}+R_{22}G_{12}\right)},
\label{eq:60z} \\
r &=& \frac{R_{11}G_{12}+R_{21}G_{22}}{R_{12}G_{11}+R_{22}G_{12}},
\label{eq:61z}
\end{eqnarray}
are slightly different in definition from those used by others~\cite{Stauffer1976,Wilemski1999} since we have off-diagonal elements $R_{12}=R_{21}$.  

Eq.~(\ref{eq:59z}) reduces to Eq.~(\ref{eq:32z})  that determines the steepest-descent direction of free-energy surface when $R_{11}=R_{22}$ and $R_{12}=R_{21}=0$.  Again, $+$ sign must be chosen when $R_{12}G_{11}+R_{22}G_{12}<0$ otherwise $-$ sign must be chosen.  Equation (\ref{eq:59z}) with $+$ sign also reduces to the well-known result~\cite{Stauffer1976,Wilemski1999} when $R_{12}=R_{21}=0$.  The $+$ sign was chosen by previous authors because they~\cite{Stauffer1976,Wilemski1999} considered the case when $G_{12}<0$.

Equation (\ref{eq:50z}) can now be written as
\begin{equation}
\frac{d^{2}\Phi}{d\zeta_{1}^{2}}=-\zeta_{1}\lambda\frac{d\Phi}{d\zeta_{1}}
\label{eq:62z}
\end{equation}
with
\begin{equation}
\lambda = L\left(\omega\right)/\Sigma\left(\omega\right).
\label{eq:63z}
\end{equation}
Since the boundary condition is given by
\begin{eqnarray}
\Phi\left(\zeta_1\right)=1,\;\;\;&\mbox{as}&\;\;\;\zeta_1\rightarrow -\infty,
\label{eq:64z} \\
\Phi\left(\zeta_1\right)=0,\;\;\;&\mbox{as}&\;\;\;\zeta_1\rightarrow \infty,
\label{eq:65z}
\end{eqnarray}
the solution of Eq.~(\ref{eq:62z}) is given by
\begin{equation}
\Phi\left(\zeta_1\right) = \frac{1}{2}{\rm erfc}\left(\sqrt{\frac{\lambda}{2}}\zeta_{1}\right),
\label{eq:67z}
\end{equation}
where ${\rm erfc}$ is the complementary error function.  Therefore, the gradient of $\Phi$ is given by
\begin{equation}
{\bm \nabla}\Phi={\bm e}_{\zeta_1}\frac{d\Phi}{d\zeta_1}
\label{eq:68z}
\end{equation}
with
\begin{equation}
{\bm e}_{\rm \zeta_1}=\cos\omega{\bm e}_{n_1}+\sin\omega{\bm e}_{n_2}
\label{eq:69z}
\end{equation}
from Eq.~(\ref{eq:35z}) and the rotation angle $\omega$ is given by Eq.~(\ref{eq:59z}).  It is easily confirmed by direct calculation that the direction of ${\bm e}_{\zeta_1}$ given by Eqs.~(\ref{eq:59z})-(\ref{eq:61z}) is in fact the direction of the eigenvector of the matrix ${\bm G}{\bm R}$ (Appendix). 

Equation (\ref{eq:67z}) indicates that the steady state cluster distribution  $f\left(n_1,n_2\right)$ is smaller than the equilibrium distribution $f_{\rm eq}\left(n_1,n_2\right)$ since $\Phi<1$ and $f=f_{\rm eq}/2$ at the saddle point.  Also,
$\Phi=f/f_{\rm eq}$ becomes a universal function along $\zeta_1$ axis.  Numerical results of vapor to liquid binary nucleation~\cite{Wyslouzil1999,Li2000} also suggest that whole contour of $\Phi=f/f_{\rm eq}$ in two-dimensional $\left(n_1,n_2\right)$ space can also be relatively insensitive to the materials considered.  Therefore the universality of Eq.~(\ref{eq:67z}) could be used as a guide to deduce the reaction coordinate on the free-energy surface and the generalized coordinates that correspond to $n_1$ and $n_2$ using various techniques of computer simulations~\cite{tenWolde1996,Pan2004,Desgranges2007,Maibaum2008}.

\subsection{\label{sec:3-3}Direction of nucleation flux $\bm J$}
Thirdly, we will derive the direction of the nucleation flux $\bm J$ following the work of Wilemski~\cite{Wilemski1999}.  Now, we rotate $\left(\Delta n_1,\Delta n_2\right)$ axis by $\phi$ and introduce a new coordinate $\left(\mu_1,\mu_2\right)$. 
Then, the nucleation flux given by Eq.~(\ref{eq:15z}) will be transformed into
\begin{equation}
{\bm J}_{\bm \mu}=f_{\rm eq}\left({\bm n}\right){\bm V}^{\rm T}\left(\phi\right){\bm R}{\bm V}\left(\phi\right){\bm\nabla}_{\bm \mu}\Phi
\label{eq:73z}
\end{equation}
similar to Eq.~(\ref{eq:40z}).

Using a procedure similar to that used in the previous subsection and Appendix A~\cite{Langer1969,Wilemski1999}, we obtain the relationship between $\phi$ and $\omega$ given by
\begin{equation}
\tan\phi=\frac{R_{12}+R_{22}\tan\omega}{R_{11}+R_{12}\tan\omega},
\label{eq:84z}
\end{equation}
which reduces to the formula derived previously~\cite{Stauffer1976,Wilemski1999} when $R_{12}=R_{21}=0$.  

Similarly, the nucleation flux $\vec{J}$ is given by
\begin{equation}
\vec{J}=J_{\mu_1}{\bm e}_{\mu_1}
\label{eq:89z}
\end{equation}
with
\begin{equation}
J_{\mu_1}=f_{\rm eq}\left({\bm n}\right)\left({\rm det}{\bm R}\right)\left[\sigma\left(\phi\right)\right]^{-1}
\sqrt{\frac{\lambda}{2\pi}}\exp\left(-\frac{1}{2}\lambda\zeta_{1}^{2}\right),
\label{eq:90z}
\end{equation}
where $\sigma\left(\phi\right)$ is given by
\begin{eqnarray}
\sigma\left(\phi\right) &=&
\left[\left(R_{11}^{2}\sin^{2}\phi+R_{22}^{2}\cos^{2}\phi\right)\right.
\nonumber \\
&&\left.-2R_{12}\left(R_{11}+R_{22}\right)\cos\phi\sin\phi
+R_{12}^{2}\right]^{1/2}.
\label{eq:87z}
\end{eqnarray}
and its direction is 
\begin{equation}
{\bm e}_{\mu_1}=\cos\phi{\bm e}_{n_1}+\sin\phi{\bm e}_{n_2}
\label{eq:91z}
\end{equation}
with the rotation angle $\phi$ given by Eq.~(\ref{eq:84z}).  

Equation (\ref{eq:84z}) tells us that the direction ${\bm v}_{\mu_1}^{\rm T}=\left(\cos\phi,\sin\phi\right)$  of the nucleation flux ${\bm e}_{\mu_1} (\propto{\bm e}_{\xi_1})$ and the direction ${\bm v}_{\zeta_1}^{\rm T}=\left(\cos\omega,\sin\omega\right)$ of the gradient of $\Phi$ (${\bm e}_{\zeta_1}$) is related through ${\bm v}_{\mu_1}\propto {\bm v}_{\xi_1}\propto {\bm R}{\bm v}_{\zeta_1}$, which again confirms the statement in the Appendix that the direction of the eigenvector of the negative eigenvalue $\lambda_1$ of the matrix ${\bm G}{\bm R}$ represents the direction of the gradient of $\Phi$ and that of the matrix ${\bm R}{\bm G}$ represents the direction of the nucleation flux $\bm J$ at the saddle point.

\subsection{\label{sec:sec3-4}Nucleation rate}
Finally, we will consider the nucleation rate.  Since the general theory was already given by Trinkaus~\cite{Trinkaus1983}, we will simply apply his theory to the two-dimensional case in this subsection.  In Appendix B, we give a detailed derivation of the formula for the nucleation rate by Trinkaus~\cite{Trinkaus1983} and the relationship between the Trinkaus' theory and Kramers-Langer-Berezhkovskii-Szabo (KLBS) theory~\cite{Kramers1940,Langer1969,Berezhkovskii2005}. 

To simplify the notation, we will introduce matrices
\begin{equation}
{\bm r}={\bm R}^{1/2}=
\begin{pmatrix}
r_{11} & r_{12} \\
r_{21} & r_{22}
\end{pmatrix}
\label{eq:92z}
\end{equation}
and 
\begin{equation}
{\bm \Gamma}={\bm r}{\bm G}{\bm r}
=
\begin{pmatrix}
\Gamma_{11} & \Gamma_{12} \\
\Gamma_{21} & \Gamma_{22}
\end{pmatrix}
\label{eq:93z}
\end{equation}
defined in Eq.~(\ref{eq:a20}).  We also introduce the rotation of coordinate ${\bm V}\left(\alpha\right)$ in Eq.~(\ref{eq:17z}) with rotation angle $\alpha$ that corresponds to the orthonormal transformation in Eq.~(\ref{eq:a27}).

Now, the nucleation flux is given as Eqs.~(\ref{eq:89z})-(\ref{eq:91z}) is also written by Eqs.~(\ref{eq:a40})  and (\ref{eq:a41}).  Therefore
\begin{equation}
\lambda\zeta_{1}^{2}=-\beta\lambda_1\xi_{1}^{2}.
\label{eq:109z}
\end{equation}
and 
\begin{equation}
\lambda=\frac{\beta\lambda_{1}}{{\rm det}{\bm R}}\left(\frac{\sigma\left(\phi\right)}{\rho\left(\phi\right)}\right)^{2}.
\label{eq:111z}
\end{equation}
since
\begin{equation}
\left|{\bm e}_{\xi_1}\right|={\rm det}{\bm R}^{1/2}/\rho\left(\phi\right)
\label{eq:106z}
\end{equation}
and
\begin{eqnarray}
\rho\left(\phi\right) &=&
\left[\left(r_{11}^{2}\sin^{2}\phi+r_{22}^{2}\cos^{2}\phi\right)\right.
\nonumber \\
&&\left.-2r_{12}\left(r_{11}+r_{22}\right)\cos\phi\sin\phi+r_{12}^{2}\right]^{1/2},
\label{eq:104z}
\end{eqnarray}
using the direction  $\phi$ of the nucleation flux.

Using the same procedure as that used in section \ref{sec:3-1}, we have one negative ($\lambda_1$) and one positive ($\lambda_2$) eigenvalues
\begin{eqnarray}
\lambda_1 &=& \left(\left(\Gamma_{11}+\Gamma_{22}\right)-\sqrt{\left(\Gamma_{11}
+\Gamma_{22}\right)^2-4{\rm det}{\bm\Gamma}}\right)/2<0, \nonumber \\
\label{eq:97z} \\
\lambda_2 &=& \left(\left(\Gamma_{11}+\Gamma_{22}\right)+\sqrt{\left(\Gamma_{11}
+\Gamma_{22}\right)^2-4{\rm det}{\bm\Gamma}}\right)/2>0, \nonumber \\
\label{eq:98z} 
\end{eqnarray}
and the rotation angle $\alpha$ is given by
\begin{equation}
\tan\alpha=\frac{\Gamma_{22}-\Gamma_{11}}{2\Gamma_{12}}
\pm\sqrt{\left(\frac{\Gamma_{22}-\Gamma_{11}}{2\Gamma_{12}}\right)^{2}+1}
\label{eq:99z}
\end{equation}
where $+$ sign must be chosen if $\Gamma_{12}<0$ otherwise $-$ sign must be chosen. 
Since, the direction of nucleation flux ${\bm e}_{\xi_1}$ in Eq.~(\ref{eq:a41}) must be parallel to ${\bm e}_{\mu_1}$ in Eq.~(\ref{eq:91z}), we have
\begin{equation}
\tan\phi=\frac{r_{12}+r_{22}\tan\alpha}{r_{11}+r_{12}\tan\alpha}
\label{eq:101z}
\end{equation}
which will be reduced to the form obtained by Wilemski~\cite{Wilemski1999} as $r_{11}=R_{11}^{1/2}$ and $r_{22}=R_{22}^{1/2}$ when $R_{12}=R_{21}=0$ ($r_{12}=r_{21}=0$). 

Finally, the nucleation rate is given by
\begin{equation}
I=f_0 e^{-\beta G^{*}}\sqrt{{\rm det}{\bm R}\frac{\left|\lambda_{1}\right|}{\lambda_2}},
\label{eq:113z}
\end{equation}
from Eq.~(\ref{eq:a48}).

In the next section, we will study the three angles $\theta$, $\omega$ and $\phi$ which represent the steepest-descent direction on the free-energy surface, gradient of $\Phi$ (commitment probability), and the nucleation flux $\bm J$.  These directions are characterized by $\theta$, $\omega$ and $\phi$ which correspond to the directions of the eigenvector of the negative eigenvalue of the matrix $\bm G$, ${\bm G}{\bm R}$ and ${\bm R}{\bm G}$, respectively~\cite{Langer1969,Berezhkovskii2005}.

\section{\label{sec:sec4}Application to the composite nucleus}

In this section, we will apply the theory of nucleation flux of binary nucleation at the saddle point developed in sec. \ref{sec:sec3} to the problem of nucleation flux of composite nucleus considered in sec. \ref{sec:sec2}.  In this case, the reaction rate matrix is non-diagonal,
\begin{equation}
{\bm R}=
\begin{pmatrix}
\kappa^{+} & -\kappa^{+} \\
-\kappa^{+} & \kappa^{+}+\alpha^{+}
\end{pmatrix}
\label{eq:114z}
\end{equation}
as given in Eq.~(\ref{eq:15z}).  Similar models with non-diagonal reaction matrix were considered by Russel~\cite{Russell1968} as a linked flux model and by Djikaev~\cite{Djikaev2002} to study the nucleation in deliquescence using the theory of Melikhov et al.~\cite{Melikhov1991}.  However the latter author used the theory~\cite{Melikhov1991} which reduces the two-dimensional problem to an approximate one-dimensional problem.  Furthermore,  only a numerical result of deliquescence path is given  and a detailed theoretical analysis is missing~\cite{Djikaev2002,Russell1968}.  Here, we will study the two-dimensional problem directly using rigorously formulated theory without resorting to an approximation.

Since, the eigenvalues of $\bm \Gamma$ are the same as those of the product ${\bm R}{\bm G}$ and ${\bm G}{\bm R}$, they can easily be obtain from Eqs.~(\ref{eq:97z}) and (\ref{eq:98z}) explicitly as
\begin{eqnarray}
\lambda_{1}&&=\frac{1}{2}\left[\left(\alpha^{+} G_{22}+\kappa^{+}\left(G_{11}+G_{22}-2G_{12}\right)\right) \right. \nonumber \\
&& -\left.\sqrt{\left(\alpha^{+} G_{22}+\kappa^{+}\left(G_{11}+G_{22}-2G_{12}\right)\right)^{2}-4\alpha^{+} k^{+}{\rm det}{\bm G}}\right], \nonumber \\
\label{eq:115z} \\
\lambda_{2}&&=\frac{1}{2}\left[\left(\alpha^{+} G_{22}+\kappa^{+}\left(G_{11}+G_{22}-2G_{12}\right)\right) \right. \nonumber \\
&& +\left.\sqrt{\left(\alpha^{+} G_{22}+\kappa^{+}\left(G_{11}+G_{22}-2G_{12}\right)\right)^{2}-4\alpha^{+} k^{+}{\rm det}{\bm G}}\right], \nonumber \\
\label{eq:116z} 
\end{eqnarray}
where $\lambda_1<0$ and $\lambda_2>0$ since ${\rm det}{\bm G}<0$ at the saddle point.  Using these eigenvalues $\lambda_1$ and $\lambda_2$, the nucleation rate is given by
\begin{equation}
I=f_0 e^{-\beta G^{*}}\sqrt{\alpha^{+} \kappa^{+}\frac{\left|\lambda_{1}\right|}{\lambda_2}},
\label{eq:117z}
\end{equation}
from Eq.~(\ref{eq:113z}) since ${\rm det}{\bm R}=\alpha^{+}\kappa^{+}$.  The nucleation rate depends not only on the rate $\alpha^{+}$ from the metastable parent phase to the metastable intermediate phase but also on the rate $\kappa^{+}$ from the metastable intermediate phase to the stable new phase.  Therefore, even when $\alpha^{+}\gg \kappa^{+}$, the nucleation problem does not reduce to one-component unary problem~\cite{Trinkaus1983,Melikhov1991,Shi1990,Zitserman1990}.  Since there is only one saddle point, a formula for the successive nucleation rate~\cite{Iwamatsu2011,Valencia2004,Valencia2006} cannot apply. 

The steepest-descent direction $\theta$ of free-energy surface is given by Eq.~(\ref{eq:32z}), which is determined solely by the free-energy surface near the critical point given by Eq.~(\ref{eq:25z}).  The direction $\omega$ of the gradient of $\Phi$, however, is given by a more complex formula in Eq.~(\ref{eq:59z}) because the effect of the anisotropy of reaction rate cannot be ignored.  Using the explicit form of reaction rate in Eq.~(\ref{eq:114z}) we have the direction $\phi$
\begin{equation}
\tan\phi=\frac{-\kappa^{+}+\left(\kappa^{+}+\alpha^{+}\right)\tan\omega}{\kappa^{+}-\kappa^{+}\tan\omega},
\label{eq:120z}
\end{equation}
once we know $\omega$ from Eqs.~(\ref{eq:59z}).

\begin{figure}
\includegraphics[width=1.0\linewidth]{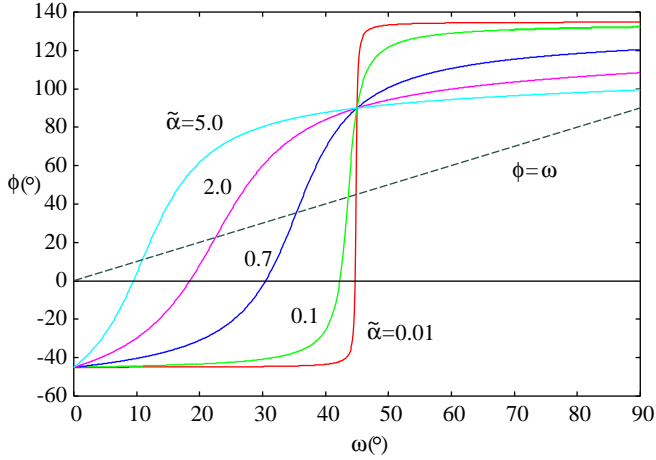}
\caption{Relation between $\omega$ and $\phi$ for various relative reaction rate $\tilde{\alpha}=\alpha^{+}/\kappa^{+}$.  The angle $\phi$ is an increasing function of $\omega$ starting from $\phi=-45^{\circ}$ at $\omega=0^{\circ}$.  The straight line indicates the line when $\phi=\omega$.  Below this line, the direction of the nucleation current is deflected from the direction of ${\rm grad}\Phi$ to the right.  }
\label{fig:2x}
\end{figure}

Figure 2 shows the relationship between $\omega$ and $\phi$ for various relative reaction rate $\tilde{\alpha}=\alpha^{+}/\kappa^{+}$.  The angle $\phi$ is an increasing function of the angle $\omega$ starting from $\phi=-45^{\circ}$ at $\omega=0^{\circ}$ from Eq,~(\ref{eq:120z}).  The direction $\phi$ of nucleation current $\bm J$ changes from right ($\phi<\omega$) to left ($\phi>\omega$) of the direction $\omega$ of the gradient of $\Phi$ as the angle $\omega$ is increased.  As the angle $\omega$ is further increased, the angle $\phi$ becomes $\phi=90^{\circ}$ when $\omega=45^{\circ}$ from Eq.~(\ref{eq:120z}) and continue to increase.  Therefore, as the direction $\omega$ of the gradient $\Phi$ increases, the nucleation flux at the saddle point turns toward the $n_2$-axis and the component $n_2$ of the intermediate metastable phase rather than $n_1$ of the stable new phase tends to grow.

When $\tilde{\alpha}<1$ ($\alpha^{+}<\kappa^{+}$), the incoming rate $\alpha^{+}$ from the metastable parent phase to the intermediate metastable phase ($n_2$) is slower than the internal transformation rate $\kappa^{+}$ from the intermediate metastable phase to the new stable phase.  Then the stable new phase ($n_1$) tends to grow faster than the intermediate metastable phase ($n_2$), and the direction $\phi$ of nucleation flux remains toward the direction of $n_1$ axis and the inequality $\phi<\omega$ follows as long as $\omega$ is not as large as $45^{\circ}$ (see Fig.~\ref{fig:2x}, $\tilde{\alpha}=0.1, 0.7$).  When  $\tilde{\alpha}>1$, the incoming rate $\alpha^{+}$ is faster than the internal transformation rate $\kappa^{+}$ and the intermediate metastable phase grows faster than the new stable phase.  Then the direction $\phi$ soon becomes larger than $\omega$ as $\omega$ is increased from 0  (see Fig.~\ref{fig:2x}, $\tilde{\alpha}=2.0, 5.0$).

In order to study the relationship and the absolute magnitude of various angles $\theta$, $\omega$, and $\phi$, we have to specify the free-energy surface $G\left({\bm n}\right)$.  To enable the qualitative idea of the composite nucleus, we will give an example to illustrate the concept and study a model system specified by the free-energy landscape
\begin{eqnarray}
G\left({\bm n}\right)&=&G^{*} \nonumber \\
&+&\frac{1}{2}G_{22}\left[\left(1+\frac{2}{\sqrt{3}}{\rm g}_{12}\right)\Delta n_{1}^{2}+\Delta n_{2}^2+2{\rm g}_{12}\Delta n_1\Delta n_2\right] \nonumber \\
\label{eq:121z}
\end{eqnarray}
with ${\rm g}_{12}<0$.  In this case $G_{11}=G_{22}\left(1+2/\sqrt{3}{\rm g}_{12}\right)$, $G_{12}=G_{22}{\rm g}_{12}$, and we find from Eq.~(\ref{eq:32z}):
\begin{equation}
\tan\theta=\frac{1}{\sqrt{3}},\;\;\;\mbox{or}\;\;\;\theta=30^{\circ}.
\label{eq:122z}
\end{equation}
In this model, only ${\rm g}_{12}<0$ is the parameter that characterizes the free-energy landscape.  In order to make the origin $(\Delta n_1, \Delta n_2)=(0,0)$ the saddle point, ${\rm det}{\bm G}$ must be negative and  ${\rm g}_{12}$ must satisfy ${\rm g}_{12}<-1/\sqrt{3}=-0.577$.

\begin{figure}[htbp]
\begin{center}
\includegraphics[width=0.75\linewidth]{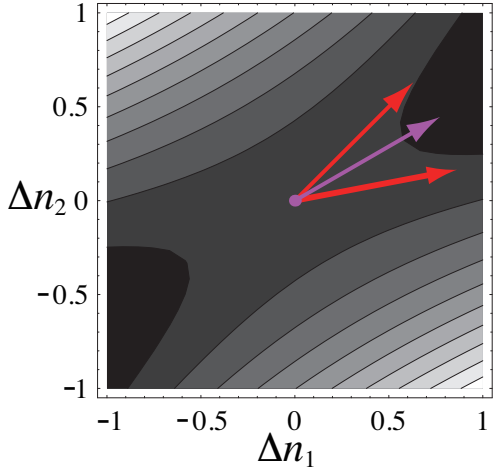}
\end{center}
\caption{
An example of the model free-energy landscape of composite nucleus when $\theta=30^{\circ}$ defined by Eq.~(\ref{eq:121z}) when ${\rm g}_{12}=-0.7$.  The origin is the saddle point shifted to $(0, 0)$. The steepest-descent direction is always fixed to $\theta=30^{\circ}$ indicated by a pink arrow from the saddle point indicated by a pink point. The direction $\omega$ of the gradient of $\Phi$ in Fig.~\ref{fig:4} is indicated by a red thin ($\tilde{\alpha}=0.0$) and a thick ($\tilde{\alpha}=7.0$) arrows. The direction $\phi$ of the nucleation flux $\bm J$ is not shown as it stays near the steepest-descent direction $\theta=30^{\circ}$ represented by a pink arrow (see also Fig.~\ref{fig:5}).  } 
\label{fig:3}
\end{figure}

In Figs.~\ref{fig:3} and \ref{fig:4}, we show contour plots of typical free-energy landscape of our model in Eq.~(\ref{eq:121z}) when ${\rm g}_{12}=-0.7$ and ${\rm g}_{12}=-5.0$.  The saddle point is located at the origin $(0, 0)$.   Although the direction of the steepest-descent is fixed to $30^{\circ}$, the saddle point and the valley that lead to the saddle becomes wider and shallower as the parameter ${\rm g}_{12}$ becomes more negative (cf. Figs.~\ref{fig:3} and \ref{fig:4}). 

\begin{figure}[htbp]
\begin{center}
\includegraphics[width=0.75\linewidth]{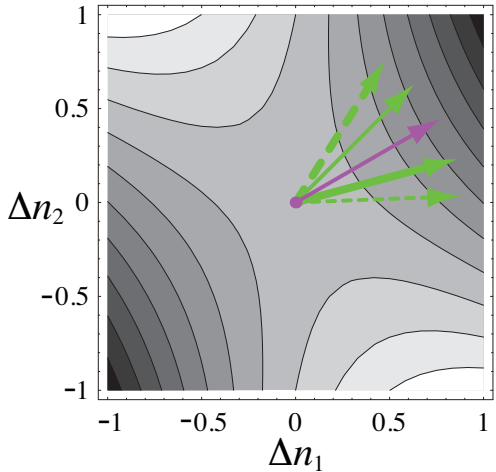}
\end{center}
\caption{
The same as Fig.~\ref{fig:3} when ${\rm g}_{12}=-5.0$.  Now the value of ${\rm g}_{12}$ becomes more negative than that in Fig.~\ref{fig:3} and the saddle point and the valley that leads to the saddle become wider. The direction $\omega$ of the gradient of $\Phi$ in Fig.~\ref{fig:5} is indicated by green thin ($\tilde{\alpha}=0.0$) and thick ($\tilde{\alpha}=7.0$) arrows of solid line. The direction $\phi$ of nucleation flux $\bm J$ is indicated by green thin ($\tilde{\alpha}=0.0$) and thick ($\tilde{\alpha}=7.0$) arrows of broken line.  } 
\label{fig:4}
\end{figure}

Figure~\ref{fig:5} shows the angles  $\omega$ and $\phi$ as the functions of the relative reaction rate $\tilde{\alpha}$.  Since we always have $R_{12}G_{11}+R_{22}G_{12}=-\kappa^{+}G_{11}+\left(\kappa^{+}+\alpha^{+}\right)G_{12}<0$ because $G_{11}>0$ and $G_{12}<0$, we took $+$ sign in Eq.~(\ref{eq:59z}).   We can easily show from Eqs.~(\ref{eq:59z}),  (\ref{eq:120z}) that
\begin{equation}
\tan\omega\rightarrow 1-\frac{G_{22}-G_{12}}{G_{11}+G_{22}-2G_{12}}\left(\frac{\alpha^{+}}{\kappa^{+}}\right),\;\;\;\mbox{as}\;\;\;\alpha^{+}\rightarrow 0^{+}.
\label{eq:123z}
\end{equation}
Therefore
\begin{equation}
\omega\rightarrow 45^{\circ},\;\;\;\mbox{as}\;\;\;\alpha^{+}\rightarrow 0^{+}
\label{eq:124z}
\end{equation}
and
\begin{equation}
\tan\phi\rightarrow \frac{G_{11}-G_{12}}{G_{22}-G_{12}},\;\;\;\mbox{as}\;\;\;\alpha^{+}\rightarrow 0^{+}
\label{eq:125z}
\end{equation}
from Eq.~(\ref{eq:120z}).  The angle $\omega$ that specifies the direction of the gradient of $\Phi$ always starts from $\omega=45^{\circ}$ at $\tilde{\alpha}=0$ irrespective of the shape of the free-energy surface $G\left({\bm n}\right)$.

\begin{figure}
\includegraphics[width=1.0\linewidth]{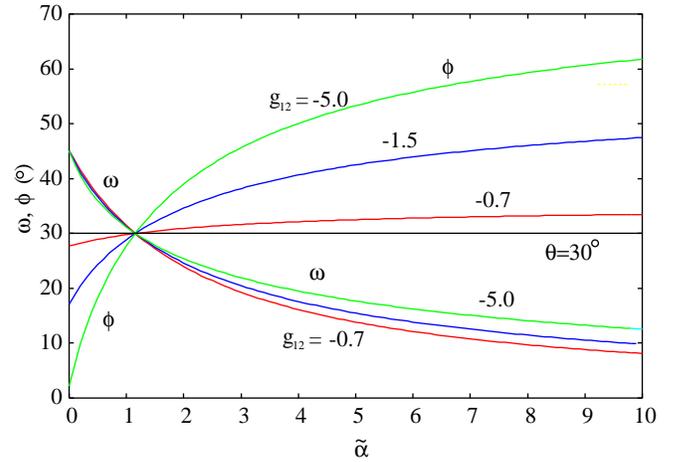}
\caption{The angle $\omega$ and $\phi$ as the functions of the relative reaction rate $\tilde{\alpha}$ for various free-energy parameters ${\rm g}_{12}$ when $\theta=30^{\circ}$. }
\label{fig:5} 
\end{figure}

As the relative reaction rate $\tilde{\alpha}$ is increased, the incoming flux from the metastable parent phase to the surrounding wetting layer of the intermediate metastable phase increases.  Then the intermediate metastable component $n_2$ of the composite nucleus increases and the nucleation flux tends toward direction of $n_2$ axis and the angle $\phi$ increases.  In Fig.~\ref{fig:5}, the angle $\omega$, which is the direction of gradient of $\Phi$, starts from $\omega=45^{\circ}$ and rotates clockwise towards $n_{1}$ axis, while the angle $\phi$, which is the direction of nucleation flux, starts from the right-hand side of the direction $\theta$ and rotates anti-clockwise and turns to the left-hand side of the direction of $\theta$ as the relative reaction rate $\tilde{\alpha}$ is increased. 

In Figs.~\ref{fig:3} and \ref{fig:4}, we show the direction of $\omega$ and $\phi$ at $\tilde{\alpha}=0.0$ and $7.0$ by thin and thick arrows using the same color as those used in Fig.~\ref{fig:5}. In Fig.~\ref{fig:4}, the arrow of solid line which represents the direction of $\omega$ of the gradient of $\Phi$ rotates clockwise, while that of broken line which represents the direction of $\phi$ of the nucleation flux $\bm J$ rotates anti-clockwise as $\tilde{\alpha}$ is increased.  It can be shown analytically that both the directions $\omega$ and $\phi$ become the same as the steepest-descent direction $\theta=\phi=\omega=30^{\circ}$ at $\tilde{\alpha}=2/\sqrt{3}$ (see Fig.~\ref{fig:5}).  When ${\rm g}_{12}=-1/\sqrt{3}$, the direction $\phi$ is fixed to the steepest-descent direction $\phi=\theta=30^{\circ}$ because the free-energy landscape becomes the corridor of valley rather than the saddle.

In Fig.~\ref{fig:7}, we show the direction $\omega$ and $\phi$ as the function of the relative reaction rate $\tilde{\alpha}$ when the steepest-descent direction is $\theta=60^{\circ}$ using the model free-energy landscape
\begin{eqnarray}
G\left({\bm n}\right)&=&G^{*} \nonumber \\
&+&\frac{1}{2}G_{22}\left[\left(1-\frac{2}{\sqrt{3}}{\rm g}_{12}\right)\Delta n_{1}^{2}+\Delta n_{2}^2+2{\rm g}_{12}\Delta n_1\Delta n_2\right]. \nonumber \\
\label{eq:126z}
\end{eqnarray}
As the relative reaction rate $\tilde{\alpha}$ is increased, the angle $\omega$ starts from $\omega=45^{\circ}$ and rotate clockwise towards $n_{1}$ axis again.  The angle $\omega$ is relatively insensitive to the parameter ${\rm g}_{12}$ of the free-energy surface not only in Fig.~\ref{fig:7} but also in Fig~\ref{fig:5}.  Several authors observed similar behaviors of $\omega$ and $\Phi$ in vapor-liquid nucleation~\cite{Wilemski1999,Li1999,Wyslouzil1999} which are relatively insensitive to the system considered.  In our case, the decreasing $\omega$ indicates that the size distribution function $f$ will be characterized mainly by the number of molecules $n_{1}$ in the final stable phase ($f\left(n_1, n_2\right)\simeq f\left(n_1\right)$) as $\tilde{\alpha}$ is increased.  The size distribution $f\left(n_1,n_2\right)$ is flat along $n_2$ axis for fixed $n_1$ and many clusters with the same $n_1$ and different $n_2$ coexist.

On the other hand, the angle $\phi$ starts from the angle $\phi>60^{\circ}$ calculated from Eq.~(\ref{eq:125z}) that is larger than the steepest-descent direction $\theta=60^{\circ}$ and continues to increase.   When ${\rm g}_{12}=-\sqrt{3}$, the direction $\phi$ is fixed again to the steepest-descent direction $\phi=\theta=60^{\circ}$ because the free-energy landscape becomes corridor of a valley.  In Fig.~\ref{fig:7} the angle $\omega$ and $\phi$ will never coincide.  As expected, the direction  $\phi$ of the nucleation flux is closer than the direction $\omega$ of the gradient of $\Phi$ to the direction $\theta$ of the steepest-descent of the free-energy surface.  Our results in Figs.~\ref{fig:5} and \ref{fig:7} clearly show that the difference between the steepest-descent direction $\theta$ and the nucleation flux direction $\phi$ depends not only on the difference of the reaction rates represented by $\tilde{\alpha}$ but also to the shape of the free-energy surface as pointed out by Wyslouzil and Wilemski~\cite{Wyslouzil1995}.

\begin{figure}[htbp]
\includegraphics[width=1.0\linewidth]{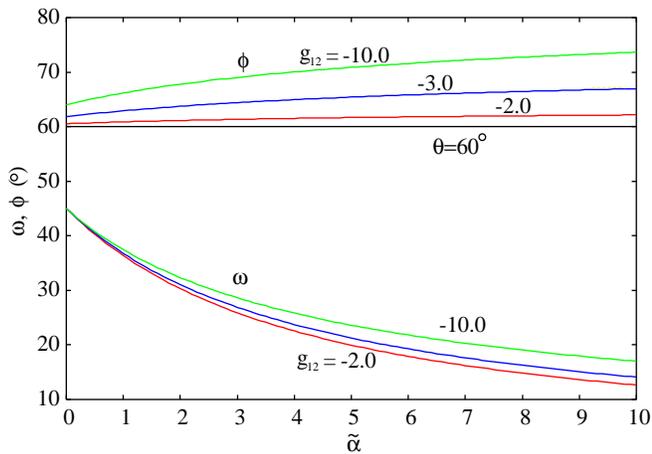}
\caption{
The same as Fig~\ref{fig:5} when  $\theta=60^{\circ}$. } 
\label{fig:7}
\end{figure}

The deviation of the direction of the nucleation flux from that of the steepest-descent direction of free-energy surface has been pointed out repeatedly in different context from condensed matter~\cite{Russell1968,Trinkaus1983, Temkin1984,Greer1990} to liquid-vapor systems~\cite{Reiss1950,Stauffer1976,Wilemski1999}.  However those authors except Trinkaus~\cite{Trinkaus1983} used various approximations, in particular, the reaction rate matrix is assumed to be diagonal.  Therefore, their conclusion remained qualitative.  In contrast, we have used the theory~\cite{Kramers1940,Langer1969,Berezhkovskii2005} which includes the non-diagonal element of the reaction rate matrix $\bm R$ to study the direction of nucleation flux.  Therefore, our result will be useful to study the nucleation flux of more complex situation and will contribute to deeper understanding of a more complex nucleation process in the future.

\section{\label{sec:sec5}Concluding remark}
In this paper, we have studied the nucleation flux of composite nucleus that consists of a stable new phase surrounded by a wetting layer of intermediate metastable phase nucleated from the metastable parents phase.  By using the results of exact formulation of nucleation flux~\cite{Kramers1940,Langer1969,Berezhkovskii2005}, we have shown that the nucleation rate is determined not only from the reaction rate from the intermediate metastable phase to the stable new phase but also from that from the metastable parent phase to the intermediate metastable phase.  We have also shown that the direction of nucleation flux deviates significantly from that of the steepest-descent direction of the free-energy surface.

Kinetics of composite nucleus has been considered by several authors in different contexts.  Kelton and coworkers~\cite{Kelton2000, Diao2008} have extended the theory of linked flux theory of nucleation by Russell~\cite{Russell1968} and studied the partitioning transition where the interplay of diffusional and interfacial processes are linked.  In this case, the nucleus consists of a core of new stable phase surrounded by a layer of pre-nucleus zone.  They~\cite{Kelton2000,Diao2008} did not show any analysis of the nucleation flux since they do not appear to be interested in the size distribution of nucleus which is directly comparable to a real experiment.  In fact, Kelton~\cite{Kelton2003} applied the theory to the oxygen precipitation in silicon and obtained a precipitation density that agrees fairly well with their experiment.  Russell~\cite{Russell1968} showed only numerically that the nucleation flux will deviate from the steepest-descent direction of free-energy surface. Very recent study by Peters~\cite{Peters2011}, however, does not use the wetting layer as a model of diffusion but considers the coupling of diffusion and nucleation directly.  His results clearly demonstrated that the nucleation pathway is influenced by the diffusion represented by along-wavelength fluctuation and also by the initial state and questioned the adequacy of the composite nucleus model. Recent studies~\cite{Philippe2011,Lutsko2012} using the dynamical density functional theory also pointed out the importance of long-wavelength fluctuation as the pre-nucleus state and questioned the composite nucleus model as a model of two-step nucleation. 

A similar composite nucleus that appears in the problem of deliquescence has been considered by Djikaev~\cite{Djikaev2002}.  In this model~\cite{Djikaev2002} the wetting layer is not merely a dense pre-nucleus zone~\cite{Kelton2000,Diao2008,Russell1968} but is rather a true thermodynamic liquid phase.  Therefore his model is closer to ours~\cite{Iwamatsu2011} since we consider the wetting layer as a metastable thermodynamic phase.  For a realistic model, he~\cite{Djikaev2002} showed only numerically that the deliquescence path, which is the direction of nucleation flux, deviates from the direction of the equilibrium path of the steepest-descent on the free-energy surface.  In contrast, our analytical analysis is not restricted to specific materials and will be useful for future extension and applications to various complex materials.

It must be noted, however, our analysis is completely confined to the steady-state process.  Therefore, transient properties~\cite{Kelton2000,Diao2008,Shi1990,Zitserman1990} and the problem of time scale~\cite{Melikhov1991,Djikaev2002} are out of scope for this study, which can only be examined numerically by solving coupled Master equation~\cite{Kelton2000,Diao2008,Wyslouzil1996}.  Also, our analysis assumed that the nucleation flux goes through the saddle point.  Saddle point avoidance~\cite{Trinkaus1983,Zitserman1990,Berezhkovskii1995,Wyslouzil1995,Wilemski1995} will be important if the anisotropy of the reaction matrix $R$ is large or the ridge between saddle point is low, which can occur at high temperatures or near the spinodal point.  In such a case, ridge-crossing rather than saddle-crossing may occur.  Then the nucleation flux will spread over the whole phase space and the picture used in this study may break down.  In such a case, the growth of composite nucleus will be expected to be more complex.

Finally, our analysis will be useful in the study of nucleation by attachment and detachment of molecules as well as the study of various barrier crossing phenomena such as chemical reactions and conformation transformation of molecules.  Newly developed various computer simulation techniques make it possible not only to explore the free-energy landscape~\cite{tenWolde1996,Desgranges2007,Maibaum2008} but also to extracts the reaction rate matrix~\cite{Ma2006,Peters2009} and identify reaction coordinate~\cite{Ma2005,Peters2006,Peters2007,Bolhuis2009,Peters2009}.  Therefore, our analysis will be useful to deduce important parameters from those numerical data extracted from simulations.  

Our analysis will also be useful even when multiple saddles exist~\cite{Iwamatsu2011,Ray1985,Wagner2001,Chen2003,Nellas2006} and multiple pathways of nucleation coexist.  In this case, the nucleation flux may fork into multiple streams and an analogy to the electric current in a circuit with parallel and series connections of conductors or resistors will be useful~\cite{Iwamatsu2012} as long as the fluxes go through those saddles.  Our results will be valid if some of those multiple saddles correspond to the composite nucleus.

\begin{acknowledgments}
This work was supported by the Grant-in-Aid for Scientific Research [Contract No.(C)22540422] from Japan Society for the Promotion of Science (JSPS) and a project for strategic advancement of research infrastructure for private universities, 2009-2013, from MEXT, Japan.  The author is grateful to one of the reviewers who pointed out importance of KLBS theory for his constructive suggestion to improve the original manuscript.

\end{acknowledgments}

\appendix
\section{Determination of the direction of gradient of $\Phi$}

In this appendix, we will determine the $\zeta$ coordinate system such that the direction of $\zeta_1$ axis corresponds to the direction of gradient of $\Phi$ using the same procedure originally developed by Langer~\cite{Langer1969} and used later by Wilemski~\cite{Wilemski1999}.  Now, we will choose the new coordinate $\zeta$ such that $\partial_{\zeta_2}\Phi=0$.  Then, 
\begin{eqnarray}
J_{\zeta_1} &=& -f_{\rm eq}\Sigma\left(\omega\right)\partial_{\zeta_1}\Phi, 
\label{eq:48z} \\
J_{\zeta_2} &=& f_{\rm eq}\Delta\left(\omega\right)\partial_{\zeta_1} \Phi,
\label{eq:49z}
\end{eqnarray}
and the steady-state condition ${\bm \nabla}_{\bm \zeta}^{\rm T}{\bm J}_{\bm \zeta}=0$ can be written as a differential equation
\begin{equation}
\Sigma\left(\omega\right)\frac{d^{2}\Phi}{d\zeta_1^{2}}+p\left(\zeta_1\right)\frac{d\Phi}{d\zeta_1}=0,
\label{eq:50z}
\end{equation}
where
\begin{equation}
p\left(\zeta_1\right)=\Sigma\left(\omega\right)\left(\frac{\partial \ln f_{\rm eq}}{\partial_{\zeta_1}}\right)_{\zeta_2}-\Delta\left(\omega\right)\left(\frac{\partial \ln f_{\rm eq}}{\partial\zeta_2}\right)_{\zeta_1}.
\label{eq:51z}
\end{equation}
Then,  using the transformation (\ref{eq:35z}), 
\begin{equation}
\ln f_{\rm eq} = \beta\left(G^{*}+\frac{1}{2}{\bm \zeta}^{\rm T}{\bm V}^{\rm T}\left(\omega\right){\bm G}{\bm V}\left(\omega\right){\bm \zeta}\right)+\mbox{constant},
\label{eq:53z}
\end{equation}
and after calculating the matrix product ${\bm V}^{\rm T}{\bm G}{\bm V}$, and the derivatives $\partial \ln f_{\rm eq}/\partial \zeta_1$ and $\partial \ln f_{\rm eq}/\partial \zeta_2$, Eq.~(\ref{eq:51z}) can be written as
\begin{equation}
p\left(\zeta_1\right)=\zeta_1 L\left(\omega\right)+\zeta_2 M\left(\omega\right),
\label{eq:56z}
\end{equation}
where
\begin{eqnarray}
L\left(\omega\right)
&=&
-\beta\Sigma\left(\omega\right)\left(G_{11}\cos^{2}\omega+G_{22}\sin^{2}\omega+G_{12}\sin 2\omega\right) \nonumber \\
&&+\beta\Xi\left(\omega\right)\left(\left(G_{22}-G_{11}\right)\sin\omega\cos\omega+G_{12}\cos 2\omega\right),
\nonumber \\
\label{eq:57z} \\
M\left(\omega\right)
&=&
-\beta\Sigma\left(\omega\right)\left(\left(G_{22}-G_{11}\right)\sin\omega\cos\omega+G_{12}\cos 2\omega\right) \nonumber \\
&&+\beta\Xi\left(\omega\right)\left(G_{22}\cos^{2}\omega+G_{11}\sin^{2}\omega-G_{12}\sin 2\omega\right).
\nonumber \\
\label{eq:58z}
\end{eqnarray}
Since it is assumed that $\Phi$ does not depend on $\zeta_2$, $\zeta_2$-dependence of $p\left(\zeta_1\right)$ in Eq.~(\ref{eq:56z}) must be eliminated.  Therefore, we have to set $M=0$ in Eq.~(\ref{eq:58z}).  Using Eqs.~(\ref{eq:44z}) and (\ref{eq:45z}) and setting $M=0$ in Eq.~(\ref{eq:58z}), we obtain the rotation angle $\omega$ that satisfies Eqs. (\ref{eq:59z})-(\ref{eq:61z}).

\section{Saddle point nucleation of multicomponent systems}

In this appendix, we  provide a summary of the theory of nucleation rate for general multicomponent systems by Trinkaus~\cite{Trinkaus1983} since his presentation is too condensed to indicate the detail of the derivation of various formulas.

The starting point is the the steady-state nucleation flux represented by a column vector $\bm J$ defined by 
\begin{equation}
{\bm J}=-{\bm R}\left({\bm n}\right)f_{\rm eq}\left({\bm n}\right){\bm \nabla}_{\bm n}\frac{f\left({\bm n}\right)}{f_{\rm eq}\left({\bm n}\right)}
\label{eq:a3}
\end{equation}
in Eq.~(\ref{eq:15z}).  Using the unit vector
\begin{equation}
{\bm e}_{\bm n}=
\begin{pmatrix}
{\bm e}_{n_1} \\
{\bm e}_{n_2} \\
\cdots \\
{\bm e}_{n_m}
\end{pmatrix}
\label{eq:a4}
\end{equation}
along the orthogonal axis $\left(n_1, n_2,\dots,n_m\right)$, we can write the nucleation current explicitly in the vector notation
\begin{equation}
\vec{J}=
J_1{\bm e}_{n_1}+J_2{\bm e}_{n_2}+\cdots+J_m{\bm e}_{n_m}={\bm J}^{\rm T}{\bm e}_{\bm n}={\bm e}_{\bm n}^{\rm T}{\bm J}.
\label{eq:a5}
\end{equation}
The equilibrium cluster distribution $f_{\rm eq}\left({\bm n}\right)$ is given by the usual Boltzmann distribution similar to Eq.~(\ref{eq:5z}).  Therefore, the flux is written as
\begin{equation}
{\bm J}=-{\bm R}\left({\bm n}\right)\exp\left(-\beta G\left({\bm n}\right)\right)  {\bm \nabla}_{\bm n}\exp\left(+\beta G\left({\bm n}\right)\right)f\left({\bm n}\right),
\label{eq:a7}
\end{equation}
which will be solved with the boundary conditions
\begin{eqnarray}
f\left(\left|{\bm n}\right|\rightarrow 0\right) &\rightarrow & f_{\rm eq}\left(\left|{\bm n}\right|\rightarrow 0\right) \label{eq:a8}, \\
f\left(\left|{\bm n}\right|\rightarrow \infty \right) &\rightarrow & 0.
\label{eq:a9}
\end{eqnarray} 

In order to study the nucleation flux in Eq.~(\ref{eq:a7}) near the saddle point, we expand free energy $G\left({\bm n}\right)$ around the saddle point ${\bm n}^{*}$ according to Eq.~(\ref{eq:25z}).  Next, in order to eliminate the anisotropy introduced by the reaction rate matrix $\bm R$, we introducing a new variable $\bm \nu$ through a linear transformation 
\begin{equation}
{\bm \nu}=\left[{\bm R}^{1/2}\right]^{-1}\Delta{\bm n},
\label{eq:a13}
\end{equation}
where ${\bm R}^{1/2}$ is the symmetric square root matrix of $\bm R$. 
Since ${\bm R}^{1/2}$ is not an orthogonal transformation, the base vectors are transformed from ${\bm e}_{\bm n}$ to ${\bm e}_{\bm \nu}$ according to
\begin{equation}
{\bm e}_{\bm \nu} = {\bm R}^{1/2}{\bm e}_{\bm n}.
\label{eq:a16}
\end{equation}
It must be noted that the new basis ${\bm e}_{\nu}$ are neither orthogonal nor normalized.

In the new coordinate system $\bm \nu$,  the steady-state nucleation flux in Eq.~(\ref{eq:a5}) is written as
\begin{eqnarray}
{\bm J} = -e^{-\beta\left(G^{*}+{\bm \nu}^{T}{\bm \Gamma}{\bm \nu}/2\right)}   {\bm \nabla}_{\bm \nu} e^{+\beta\left(G^{*}+{\bm \nu}^{T}{\bm \Gamma}{\bm \nu}/2\right)} 
f\left({\bm \nu}\right)
\nonumber \\
\label{eq:a22}
\end{eqnarray}
from Eq.~(\ref{eq:a7}), where 
\begin{equation}
{\bm \Gamma}={\bm R}^{1/2}{\bm G}{\bm R}^{1/2}.
\label{eq:a20}
\end{equation}

Since the matrix ${\bm \Gamma}$ defined in Eq.~(\ref{eq:a20}) is a symmetric matrix because ${\bm G}$ and ${\bm R}^{1/2}$ are symmetric, we can diagonalize ${\bm \Gamma}$ using an orthogonal matrix ${\bm V}$ as
\begin{equation}
{\bm V}^{\rm T}{\bm \Gamma}{\bm V}={\bm \Lambda},
\label{eq:a23}
\end{equation}
where $\bm \Lambda$ is a diagonal matrix of the eigenvalues $\lambda_{i}$ given by
\begin{equation}
{\bm \Gamma}{\bm v}_{i}=\lambda_{i}{\bm v}_{i},\;\;\;i=1,\dots, n,
\label{eq:a25}
\end{equation}
where ${\bm v}_i$ is the normalized (column) eigenvectors.  We choose $\lambda_1<0$ as a single negative eigenvalues. Other eigenvalues $\lambda_2,\dots,\lambda_m$ are positive since the point ${\bm n}^{*}$ is the saddle point and ${\rm det}{\bm G}<0$. The orthogonal (orthonormal) matrix $\bm V$ is given by
\begin{equation}
{\bm V}=\left({\bm v}_1,{\bm v}_2, \dots, {\bm v}_m\right)
\label{eq:a26}
\end{equation}
using the (normalized) eigenvectors ${\bm v}_i$.

Using the this orthogonal transformation $\bm V$ from $\bm \nu$ to $\bm \xi$ by
\begin{equation}
{\bm\xi}={\bm V}^{\rm T}{\bm \nu},
\label{eq:a27}
\end{equation}
which corresponds to the rotation of axis
\begin{equation}
{\bm e}_{\bm \xi}={\bm V}^{\rm T}{\bm e}_{\bm \nu},
\label{eq:a28}
\end{equation}
we can diagonalize matrix $\bm \Gamma$.
Then, the steady-state nucleation flux is given by
\begin{eqnarray}
{\bm J} = -e^{-\beta\left(G^{*}+ {\bm \xi}^{T}{\bm \Lambda}{\bm \xi}/2\right)}  {\bm \nabla}_{\bm \xi} e^{\beta\left(G^{*}+ {\bm \xi}^{T}{\bm \Lambda}{\bm \xi}/2\right)} f\left({\bm \xi}\right), \nonumber \\
\label{eq:a34}
\end{eqnarray}
which must be a time-independent constant.

It is reasonable to assume that the nucleation flux goes to the direction ${\bm e}_{\xi_1}$ that corresponds to the negative eigenvalue $\lambda_1$ of $\bm \Gamma$ in Eq.~(\ref{eq:a25}).  Then, the multi-dimensional problem in Eq.~(\ref{eq:a34}) reduces to the one dimensional problem
\begin{eqnarray}
J_{\xi_1} = -e^{-\beta\left(G^{*}+ {\bm \xi}^{T}{\bm \Lambda}{\bm \xi}/2\right)} \frac{\partial}{\partial \xi_1} e^{\beta\left(G^{*}+ {\bm \xi}^{T}{\bm \Lambda}{\bm \xi}/2\right)} f\left({\bm \xi}\right), \nonumber \\
\label{eq:a35}
\end{eqnarray}
which can be integrated by $\xi_1$.  Using the boundary condition Eqs.~(\ref{eq:a8}) and (\ref{eq:a9}) and the expansion (\ref{eq:25z}) around the saddle point, we arrive at the equation
\begin{equation}
J_{\xi_1}=f_{\rm eq}\left({\bm n}\right)\sqrt{\frac{\beta\left|\lambda_1\right|}{2\pi}}\exp\left[\frac{\beta}{2}\lambda_{1}\xi_{1}^{2}\right]
\label{eq:a40}
\end{equation}
and
\begin{equation}
\vec{J}=J_{\xi_1}{\bm e}_{\xi_1}.
\label{eq:a41}
\end{equation}
These results were originally derived by Trinkaus~\cite{Trinkaus1983}.  We should note that the basis vector ${\bm e}_{\xi_1}$ is not normalized ($\left|{\bm e}_{\xi_1}\right|\neq 1$).  Therefore its magnitude contribute to the net magnitude of nucleation flux $\vec{J}$ in Eq.~(\ref{eq:a41}).

The nucleation rate $I$ can be calculated by integrating the nucleation flux $J_{\xi_1}$ in $m-1$ dimensional hypersurface in $\bm n$, which can be transformed into the integration by $\xi_2,\dots,\xi_m$:
\begin{equation}
I = \int_{-\infty}^{\infty}\cdots\int_{-\infty}^{\infty} J_{\xi_1}\left({\bm \xi}\right) {\rm det}{\bm R}^{1/2}d\xi_2 d\xi_3 \cdots d\xi_m,
\label{eq:a46}
\end{equation}
where ${\rm det}{\bm R}^{1/2}$ is the Jacobian to preserve the volume element.  Using Eq.~(\ref{eq:a40}) in Eq.~(\ref{eq:a46}), we find
\begin{equation}
I =
\frac{\beta\left|\lambda_1\right|}{2\pi}{\rm det}{\bm R}^{1/2}f_{0}e^{-\beta G^{*}}
\sqrt{\left|\prod_{i=1}^{m}\frac{2\pi}{\beta\lambda_{i}}\right|}.
\label{eq:a48}
\end{equation}
Since
\begin{eqnarray}
\prod_{i=1}^{m}\lambda_{i} &= &{\rm det}{\bm \Lambda}
={\rm det}{{\bm V}^{\rm T}{\bm \Gamma}{\bm V}}
={\rm det}{{\bm V}^{\rm T}{\bm R}^{1/2}{\bm G}{\bm R}^{1/2}{\bm V}}
\nonumber \\
&=&\left({\rm det}{\bm R}^{1/2}\right)^{2}{\rm det}{\bm G},
\label{eq:a49}
\end{eqnarray}
we can rewrite Eq.~(\ref{eq:a48}) in the form
\begin{equation}
I=\frac{\beta\left|\lambda_1\right|/2\pi}{\sqrt{{\rm det}\left[\beta {\bm G}/2\pi \right]}} f_0 e^{-\beta G^{*}}.
\label{eq:a50}
\end{equation}
derived originally by Langer~\cite{Langer1969} and re-derived later by Trinkaus~\cite{Trinkaus1983}.

Finally, we will note the connection to the more general KLBS theory~\cite{Kramers1940,Langer1969,Berezhkovskii2005}.
It is easy to show from Eqs.~(\ref{eq:a20}) and (\ref{eq:a25}) that
the direction ${\bm e}_{\xi_1}={\bm v}_{1}^{\rm T}{\bm e}_{\bm \nu}=\left({\bm R}^{1/2}{\bm v}_{1}\right)^{\rm T}{\bm e}_{\bm n}$ of the nucleation flux is the direction of the eigenvector of the negative eigenvalue $\lambda_1$ of the matrix ${\bm R}{\bm G}$.  Therefore, the eigenvector ${\bf f}_{+}$ defined by Berezhkovskii and Szabo~\cite{Berezhkovskii2005} represents the direction of nucleation flux.

Similarly, it is easy to show from Eqs.~(\ref{eq:a20}) and (\ref{eq:a25}) that the direction of the eigenvector of the negative eigenvalue $\lambda_1$ of the matrix ${\bm G}{\bm R}$ represents the direction of the gradient of the stochastic separatrix defined by $\Phi=1/2$~\cite{Berezhkovskii2005,Rhee2005,Ma2006}.  Therefore, the eigenvector ${\bf e}_{+}$ defined by Berezhkovskii and Szabo~\cite{Berezhkovskii2005} represents the direction ${\bm e}_{\zeta_1}$ of  $\nabla\Phi$ at the saddle point in Eqs.~(\ref{eq:68z})  and (\ref{eq:69z}) in our nucleation problem.


\end{document}